\documentclass[twocolumn,showpacs,preprintnumbers,amsmath,amssymb,prl,
superscriptaddress]{revtex4}

\usepackage{graphicx}% Include figure files
\usepackage{dcolumn}% Align table columns on decimal point
\usepackage{bm}% bold math

\newcommand{\qsq}{Q^{2}}

\begin{document}

%\preprint{APS/123-QED}

\title{New Measurements of the Transverse Beam Asymmetry for Elastic
Electron Scattering from Selected Nuclei}

\affiliation{Argonne National Laboratory, Argonne, Illinois 60439, USA} %ANL
\affiliation{Cairo University, Giza 12613, Egypt} 
\affiliation{California State University, Los Angeles, Los Angeles, California 90032, USA} %CULA
\affiliation{Carnegie Mellon University, Pittsburgh, Pennsylvania 15213, USA} %CMU  
\affiliation{CEA Saclay, DAPNIA/SPhN, F-91191 Gif-sur-Yvette, France}
\affiliation{China Institute of Atomic Energy, Beijing 102413, China}
\affiliation{Clermont Universit\'e, Universit\'e Blaise Pascal, CNRS/IN2P3,
Laboratoire de Physique Corpusculaire, FR-63000 Clermont-Ferrand, France}
\affiliation{College of William and Mary, Williamsburg, Virginia 23187, USA} %W&M
\affiliation{Christopher Newport University, Newport News, Virginia 23606, USA} %CNU
\affiliation{Duke University, TUNL, Durham, North Carolina, 27706, USA}
\affiliation{Florida International University, Miami, Florida 33199, USA} %FIU
\affiliation{The George Washington University, Washington DC 20052, USA} %GWU
\affiliation{Hampton University, Hampton, Virginia 23668, USA} %HU
\affiliation{Harvard University, Cambridge, Massachusetts 02138, USA} %Harvard 
\affiliation{INFN, Sezione di Bari and University of Bari, I-70126 Bari, Italy} %INFN/Bari
\affiliation{INFN, Sezione di Catania and University of Catania, I-95123 Catania, Italy} %INFN/Catania
\affiliation{INFN, Sezione di Roma, I-00161 Rome, Italy} %INFN Roma
\affiliation{INFN, Sezione di Roma, gruppo Sanit\`a, I-00161 Rome, Italy} %INFN/Rome
\affiliation{Indiana University, Bloomington, Indiana 47405, USA} %IUB
\affiliation{Institut Jo\v zef Stefan, 3000 SI-1001 Ljubljana, Slovenia}
\affiliation{Kent State University, Kent, Ohio 44242, USA}
\affiliation{Kharkov Institute of Physics and Technology, Kharkov 61108, Ukraine} 
\affiliation{LPSC, Universit\'e Joseph Fourier, CNRS/IN2P3, INPG, Grenoble , France}
\affiliation{Lawrence Berkeley National Laboratory, Berkelery, California 94720, USA} %LBNL
\affiliation{Longwood University, Farmville, Virginia 23909, USA} %LU 
\affiliation{Los Alamos National Laboratory, Los Alamos, New Mexico 87545, USA} %LANL
\affiliation{Massachusetts Institute of Technology, Cambridge, Massachusetts 02139, USA} %MIT
\affiliation{Mississippi State University, Mississippi State, Mississippi 39762, USA} %UMiss
\affiliation{Ohio University, Athens, Ohio 45701, USA} %OU
\affiliation{Old Dominion University, Norfolk, Virginia 23529, USA} %ODU
\affiliation{Rensselaer Polytechnic Institute, Troy, New York 12180, USA} %RPI
\affiliation{Rutgers University, The State University of New Jersey, New Brunswick, New Jersey 08901, USA} %SUNJ 
\affiliation{Seoul National University, Seoul 151-742, South Korea} %Seoul
\affiliation{Smith College, Northampton, Massachusetts 01063, USA} %Smith
\affiliation{Syracuse University, Syracuse, New York 13244, USA} %SU
\affiliation{Tel Aviv University, P.O. Box 39040, Tel-Aviv 69978, Israel} %TAU
\affiliation{Temple University, Philadelphia, Pennsylvania  19122, USA} %TU   
\affiliation{Thomas Jefferson National Accelerator Facility, Newport News, Virginia 23606, USA}  %JLab
\affiliation{University of Chicago, Chicago, Illinois, 60637, USA}
\affiliation{University of Illinois, Urbana, Illinois, 61801, USA}
\affiliation{University of Kentucky, Lexington, Kentucky 40506, USA}
\affiliation{University of Maryland, College Park, Maryland, 20742, USA}
\affiliation{University of Massachusetts Amherst, Amherst, Massachusetts  01003, USA} %UMass
\affiliation{University of New Hampshire, Durham, New Hampshire 03824, USA} %UNH
\affiliation{University of Science and Technology of China, Hefei, Anhui 230026, P.R. China} %USTC 
\affiliation{University of Virginia, Charlottesville, Virginia  22903, USA} %UVa
\affiliation{Virginia Polytechnic Institute and State University, Blacksburg, Virginia  24061, USA} %VT 
\affiliation{Yerevan Physics Institute, Yerevan, Armenia} %Yerevan
\author{S.~Abrahamyan}\affiliation{Yerevan Physics Institute, Yerevan, Armenia} %Yerevan
\author{A. Acha}\affiliation{Florida International University, Miami, Florida 33199, USA} %FIU
\author{A. Afanasev}\affiliation{The George Washington University, Washington DC 20052, USA} %GWU
\author{Z.~Ahmed}\affiliation{Syracuse University, Syracuse, New York 13244, USA} % SU
\author{H.~Albataineh}\affiliation{Clermont Universit\'e, Universit\'e Blaise Pascal, CNRS/IN2P3,
Laboratoire de Physique Corpusculaire, FR-63000 Clermont-Ferrand, France}
\author{K.~Aniol}\affiliation{California State University, Los Angeles, Los Angeles, California  90032, USA} %CULA
\author{D.~S.~Armstrong}\affiliation{College of William and Mary, Williamsburg, Virginia  23187, USA} %W&M
\author{W.~Armstrong}\affiliation{Temple University, Philadelphia, Pennsylvania  19122, USA} %TU 
\author{J. Arrington}\affiliation{Argonne National Laboratory, Argonne, Illinois 60439, USA} %ANL
\author{T.~Averett}\affiliation{College of William and Mary, Williamsburg, Virginia  23187, USA} %W&M
% ...
%USTC 
\collaboration{HAPPEX and PREX Collaborations}

\date{\today}

\begin{abstract}

We have measured the beam-normal single-spin 
asymmetry $A_{\rm n}$ in the elastic scattering of 1-3 GeV transversely polarized 
electrons from $^1$H and for the first time from $^4$He, $^{12}$C, and $^{208}$Pb.  
For  $^1$H, $^4$He and $^{12}$C, the measurements are in agreement 
with calculations that relate $A_{\rm n}$ to the imaginary part of the 
two-photon exchange amplitude including inelastic 
intermediate states.  Surprisingly, the $^{208}$Pb result is significantly smaller
than the corresponding prediction using the same formalism. These results suggest that a systematic set of new 
$A_{\rm n}$ measurements might emerge as a new and sensitive probe of the structure 
of heavy nuclei.

\end{abstract}

\maketitle

Traditionally, fixed-target electron scattering has been analyzed in terms
of the one-boson (photon or $Z$) exchange approximation.  For scattering off heavy nuclei, 
distorted waves, based on solutions to the Dirac equation in the strong electric field 
of the nucleus, are also required to describe the data.
Recently, the inclusion of the exchange of one or more additional photons has been 
necessary for the interpretation of precision data.  
The electric form factor $G_{\rm E}^{\rm p}$ extracted in elastic electron-proton scattering using two different
techniques, Rosenbluth separation and polarization observables, were inconsistent~\cite{Perdrisat:2006hj,ArringtonJPG07,Arrington2003}. The latter should be less sensitive to higher order electromagnetic effects and  
calculations including two-photon exchange provide a 
plausible explanation for the difference~\cite{Afanasev:2005mp,Carlson07,ArringtonPRC76,BMT2005,JBM2011}.  
Another example  is corrections to the parity-violating
asymmetry $A_{\rm PV}$ in the same process, which
provides a measurement of the weak charge of the
proton and serves as a sensitive test of the electroweak theory.  
For interpreting $A_{\rm PV}$, $\gamma$-$Z$ box diagrams 
are important~\cite{Gorchtein:2011mz}, as well as two-photon exchange
~\cite{Afanasev05}.  Theoretical calculations of two photon exchange processes are difficult because an integral over all off-shell proton intermediate state contributions must be made.

The effect of the extra boson is relatively small on the measured 
cross section or asymmetry for the above examples.  
On the other hand, the beam-normal spin asymmetry $A_{\rm n}$ for elastic electron
scattering at GeV energies is dominated by two (or more) $\gamma$ exchange.  
Several measurements of $A_{\rm n}$ at GeV energies for the proton 
have been 
reported \cite{Wells:2000rx,Maas:2004pd,Armstrong:2007vm,Androic:2011rh}.
Several theoretical papers report
computed values of $A_{\rm n}$ that are in 
qualitative agreement with the 
data when they include the effects of inelastic
intermediate hadronic states ~\cite{Afanasev:2004pu,Pasquini:2005yh,Pasquini:2004pv,Afanasev:2007ar}.

The beam-normal, or transverse asymmetry, $A_{\rm n}$ is a direct probe of higher-order photon exchange as time-reversal symmetry dictates that $A_{\rm n}$ is zero at first Born approximation.  
%Furthermore, measuring $A_n$ from nuclei over a range of charge Z should probe the interplay of dispersion %corrections and Z-dependent Coulomb distortion effects.  
Afanasev {\it et al.}~\cite{Afanasev:2005mp} and 
Gorchtein and Horowitz ~\cite{Gorchtein:2008dy} have
calculated $A_{\rm n}$, in a two-photon exchange approximation, but including a full
range of intermediate excited states. 
Gorchtein and Horowitz predict that 
$A_{\rm n}$ scales roughly as the ratio of mass number $A$ to $Z$, and 
is not strongly $Z$-dependent.  In
contrast Cooper {\it et al.}~\cite{Cooper:2005sk} 
calculate Coulomb distortion effects and work to all orders in photon
exchanges by numerically solving the Dirac equation.  However, they only
consider the elastic intermediate state.  They find that elastic
intermediate state contributions, while in general small, increase
strongly with $Z$.  

To predict $A_{\rm n}$ for nuclear targets, Afanasev used a unitarity-based 
model \cite{Afanasev:2007ar} with the total photoproduction cross section
and the Compton slope as input; his prediction for ${}^4$He is
consistent with the value of $A_{\rm n}$ reported in this paper.
However, there is not yet a calculation of
$A_{\rm n}$ that includes both Coulomb distortion effects and a full range of excited
intermediate states. Measuring $A_{\rm n}$ as a function of $Z$ might
reveal the role of dispersion effects relative to Coulomb distortions and motivate more detailed calculations.
To this end, in this Letter, we report data on the beam-normal 
spin-asymmetry $A_{\rm n}$
 on the targets $^1$H, $^4$He, $^{12}$C, and $^{208}$Pb.  

To observe the beam-normal single-spin asymmetry, the electron beam spin vector 
$\vec P_{\rm e}$ must have a component normal 
to the scattering plane defined by the unit vector $\hat k$ perpendicular to the plane,
where $\hat k = \vec k / |k|$; $\vec k =\vec k_{\rm e}\times\vec k_{\rm out}$, with  
$\vec k_{\rm e}$ and $\vec k_{\rm out}$ are respectively the incident and scattered electron momenta.
The measured beam-normal single-spin asymmetry is then defined as 
$A^{\rm m}_{\rm n}=(\sigma_\uparrow-\sigma_\downarrow)/(\sigma_\uparrow+\sigma_\downarrow)$
 where $\sigma_{\uparrow(\downarrow)}$ is the cross section for beam electron
spin parallel(anti-parallel) to $\hat k$.  The measured asymmetry $A_{\rm n}^{\rm m}$
is related to $A_{\rm n}$ by
\begin{equation}
A_{\rm n}^{\rm m}\hskip 0.05in=\hskip 0.05in A_{\rm n}\hskip 0.05in\vec P_{\rm e}\cdot\hat k,
\label{eq:An}
\end{equation}
where  $\phi$ is the angle between $\hat k$ and $\vec P_{\rm e}$: cos $\phi = \vec P_{\rm e}\cdot\hat k / |P_{\rm e}|$.

The measurements were carried out in Hall A at the Thomas Jefferson National 
Accelerator Facility.   The data were obtained as a part of a study of 
systematic uncertainties for three experiments designed
to measure $A_{\rm PV}$  in elastic electron scattering, since 
$A_{\rm n}$ can contribute to the extracted $A_{\rm PV}$ if the beam polarization 
has a transverse component and the apparatus lacks perfect symmetry.

The data were obtained in 2004 for the $^1$H and $^4$He targets where the primary goal was  
to measure $A_{\rm PV}$ in order to determine the
strange form factors in the nucleon~\cite{Acha:2006my,Aniol:2005zf}. 
The $^{12}$C and $^{208}$Pb data were obtained in 2010 where the goal 
was to determine the radius of the distribution 
of neutrons \cite{PREX,bigprex}. 
The kinematics for each target is given in Table~\ref{table:kin}:  the central acceptance angle of the spectrometers $\theta$,  the beam energy $E_{\rm b}$, the acceptance-averaged 4-momentum transfer $Q^2$, and 
the average accepted $\cos\phi$ (Eqn.~\ref{eq:An}).  The uncertainties in $Q^2$ were 1\% for the $^1$H 
and ${}^4$He
data and 1.3\% for the ${}^{12}$C and ${}^{208}$Pb data \cite{Acha:2006my,Aniol:2005zf,PREX}.
\begin{table}
\begin{tabular}{|c|c|c|c|c|}\hline
&&&&\\
  Target  &       H &    $^4$He &     $^{12}$C   &    $^{208}$Pb \\
\hline
$\theta$ & 6$^\circ$ & 6$^\circ$ & 5$^\circ$  &  5$^\circ$ \\
$Q^2$ (GeV$^2$) &    0.0989     & 0.0773    & 0.00984       & 0.00881 \\
$E_{\rm b}$(GeV) & 3.026     &   2.750   &    1.063    &   1.063  \\
$\langle\cos\phi\rangle$ & 0.968 & 0.967 & 0.963  & 0.967 \\
%$A_n$(ppm)&$-6.58\pm1.49$&$-13.5\pm1.39$&$-6.49\pm0.36\pm0.11$
%&$0.28\pm0.21\pm0.14$ \\
\hline 

\end{tabular}
\caption{Kinematic values for the various targets.}
\label{table:kin}
\end{table}

All of the targets except $^{12}$C were cooled with helium
gas at about 20K.  The LH$_2$ and high pressure He targets featured rapid 
vertical flow
of the fluid. In addition, the beam was rastered over a 4~mm~$\times$~4~mm square for
all targets.  The 0.55~mm thick isotopically pure $^{208}$Pb target 
was sandwiched between two 150~$\mu$m diamond foils, and the edges were cooled 
with the cold helium.  
Electrons elastically scattered from the targets
were focused onto detectors in the focal plane of the Hall A High Resolution
Spectrometers \cite{HANIM}.  The transverse asymmetry is modulated by the sine of the azimuthal electron scattering angle
and so the electron polarization was set vertical. 
This ensured that the acceptance of the two spectrometers, which 
are symmetrically placed to accept horizontally scattered events, contained the maximum and minimum of the asymmetry.  
The momentum resolution of the spectrometers ensured that essentially only elastic events were accepted.

To measure the asymmetry, Cherenkov light was produced 
in a radiator and collected by a PMT, whose output 
sent to an analog-to-digital converter (ADC) and integrated 
 over a fixed time period of constant helicity.  
These detectors had to withstand the radiation damage caused by the high
signal flux and also provide a uniform response to the electrons so that
integrating the signals did not increase fluctuations.
For the ${}^{208}$Pb and ${}^{12}$C data, each spectrometer had two 3.5 cm by 14 cm quartz detectors 
oriented at
45$^\circ$ to the direction of the electrons in the spectrometer,
one in front that was 5 mm thick and one behind that was 1 cm thick.
For the $^1$H and $^{4}$He data, a five-layer sandwich of quartz and brass provided
sufficient energy resolution.

The electron beam originated from a GaAs photocathode illuminated by 
circularly polarized light~\cite{Sinclair2007}.
By reversing the sign of the laser circular 
polarization, the direction of the spin at the target could be 
reversed rapidly \cite{Paschke:2007zz}.
A half-wave ($\lambda$/2) plate was periodically inserted into the 
laser optical path which passively reversed the
sign of the electron beam polarization. 
Roughly equal statistics were thus 
accumulated with opposite signs for the measured asymmetry, which suppressed 
many systematic effects.  
The direction of the polarization could be
controlled by a Wien filter and solenoidal lenses
near the injector \cite{GramesWien2011}.  The accelerated beam was 
directed into Hall A, where its intensity, energy and trajectory on 
target were inferred from the response of several monitoring devices.

Each period of constant spin direction is referred to as a ``window''.
The beam monitors, target, detector components and 
electronics were designed so that
the fluctuations in the fractional difference in the PMT response between
a pair of successive windows were
dominated by scattered electron counting statistics.
To keep spurious beam-induced asymmetries under
 control at well below the ppm level, 
careful attention was given to the design and configuration of the laser 
optics leading to the photocathode \cite{Paschke:2007zz}.

The spin-reversal rate was 
30 Hz for the $^1$H and ${}^4$He data and 
240 Hz for the ${}^{12}$C and ${}^{208}$Pb data.
The integrated response of each detector PMT and beam monitor
was digitized and recorded for each window.
In the 30 Hz case, the raw spin-direction asymmetry $A_{\rm raw}$
in each spectrometer arm was computed from the the detector response
normalized to the beam intensity 
for each window pair. 
At the faster reversal, quadruplets of windows with either 
of the patterns $+--+$ or $-++-$ were used to suppress
the significant 60 Hz line noise.  In either case, the sequence of these
patterns was chosen with a pseudorandom number generator.

Loose requirements were imposed on beam quality,
removing periods of beam intensity, position, or energy
instability, removing about 25\%\ of the total data sample. 
No spin-direction-dependent cuts were applied.
Since we measure the difference between two horizontal detectors,
the dominant source of noise due to the beam arose
from position fluctuations in the horizontal direction, which change the
acceptance of the spectrometers in opposite directions.  
Noise in the beam energy or current largely cancels.
In contrast, a measurement of the sum of the detectors in the
$A_{\rm PV}$ case is relatively more sensitive to beam energy or current
fluctuations and less to the beam position.

As explained in detail in \cite{Acha:2006my,Aniol:2005zf,PREX},
the window-to-window differences in the asymmetry from beam jitter were reduced 
by using the correlations to beam position differences from precision
beam position monitors, $\Delta x_{\rm i}$ by defining a correction $A_{\rm beam} = \sum c_{\rm i}\Delta x_{\rm i}$.
The $c_{\rm i}$ were measured several times each hour
from calibration data where the beam was modulated
using steering coils and an accelerating cavity.
The largest $c_{\rm i}$ was for 
${}^{208}$Pb and was on the order of 50 ppb/nm.
The spread in the resulting $A_{\rm n}^{\rm m}=A_{\rm raw}-A_{\rm beam}$ was 
 observed to be dominated by counting statistics.
For example, for ${}^{208}$Pb, which had the highest rate 
and hence the smallest statistical uncertainty for a 
window, this spread corresponded to a rate of about
1 GHz at beam current of 70 $\mu$A.
About one day was spent at each $\lambda$/2
setting on each target.
\begin{figure}
\includegraphics[width=1.0\columnwidth]{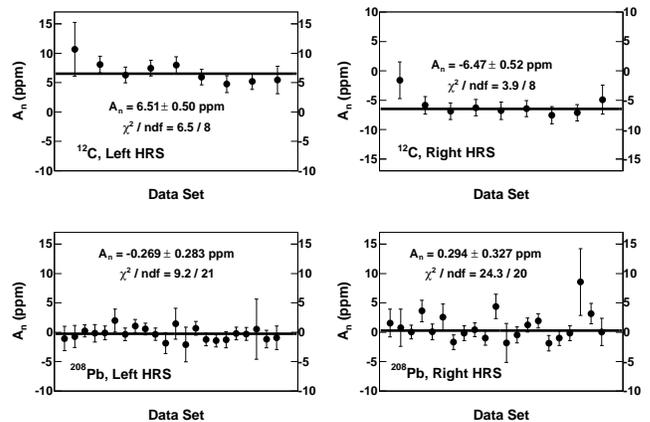}
\caption{Plots of the asymmetries
for carbon and lead.
Top row shows for the $^{12}$C target the left-HRS and right-HRS
asymmetries in the left and right panels, respectively.  
Bottom row shows the same sequence for the ${}^{208}$Pb target.
The data have been sign-corrected for the $\lambda$/2 plate 
insertions.}
\label{fig:CPb}
\end{figure}

The values of $A_{\rm n}^{\rm m}$ were consistent from run-to-run as shown in
Fig.~\ref{fig:CPb}.  The asymmetries in each spectrometer arm were of opposite sign
as expected ($\hat k$ in Eq.~\ref{eq:An} reverses sign).
After correcting for the the $\lambda$/2
reversals, the magnitudes of $A_{\rm n}^{\rm m}$ are consistent within statistical uncertainties. 
The reduced $\chi^2$ for a constant fit to the $A_{\rm n}^{\rm m}$
runs is close to one for each target type.
The average $A_{\rm beam}$ corrections  were negligible.
The physics asymmetry $A_{\rm n}$ is calculated from $A_{\rm n}^{\rm m}$ by
correcting for the beam polarization $P_{\rm e}$, the average value of $\cos\phi$ as
given in Table \ref{table:kin}, and the background subtractions
from the Al windows in the LH$_2$ and ${}^4$He targets,
and the diamond surrounding the lead foil.

Nonlinearity in the PMT response was limited to 1\%\ in bench tests that 
mimicked running conditions. The total
relative nonlinearity between the PMT response and those of the beam intensity
monitors was limited to 1.5\% by studies. 
An acceptance correction accounted for the 
non-linear dependence of the asymmetry with $\qsq$. 
A significant systematic uncertainty in $\langle\qsq\rangle$ is in the 
determination of the absolute scale of $\theta_{\rm lab}$. A
nuclear recoil technique with a dedicated calibration run using a water
cell target~\cite{Aniol:2005zf} was used to set a scale uncertainty 
on $\langle\qsq\rangle$ of $<0.2\%$.

Beam polarization measurements ($P_{\rm e}$ in Eq. \ref{eq:An})
were made during the
runs for the four nuclei.
The beam polarization was inferred from longitudinal polarization measurements
taken before and after the transverse polarization data taking.  
A solenoid was used to control the orientation of the polarization between 
the longitudinal and transverse (vertical) direction.
The polarization was verified to be purely vertical to
 within $\pm 2^\circ$ with a Mott polarimeter 
located in an injector 5 MeV extraction line.
The vertical component
of the polarization set at the injector is conserved after passing
through the CEBAF accelerator, a result of accelerating and
transporting the polarized beam in planes flat with respect to one
another.  The extent to which the beam-spin tune degrades the
vertical polarization orientation in CEBAF has been studied and
determined to be $\pm 1^\circ$~\cite{grames_thesis}.
A small longitudinal component of the electron spin introduces a 
negligible parity-violating contribution to the measured asymmetry.  
%The lead target parity-violating asymmetry was measured during the same 
%run period to be 0.5936 $\pm$ 0.0504 ppm~\cite{PREX}, and a 
%Standard Model value of 0.817 $\pm$ 0.041 ppm was used for carbon.
\begin{table}
\begin{tabular}{|c|c|c|c|c|}\hline
&&&&\\
 Target   &       $^1$H &    $^4$He  &     $^{12}$C   &    $^{208}$Pb \\
\hline
False asymmetry &  0.14 &  0.11 &       0.02    &      0.12 \\
Beam polarization &  0.21 &  0.33 &       0.08    &      0.003 \\
Linearity &         0.07 &  0.15 &       0.06    &      0.004 \\
Target Windows &    0.06 &  0.12 &       0.00    &      0.062 \\\hline
{\bf Total Systematic }        &    0.27 &  0.41 &       0.10    &      0.14 \\
{\bf Statistical  }       &    1.52 &  1.39 &       0.36    &      0.21 \\
\hline 
\end{tabular}
\caption{$A_{\rm PV}$ uncertainty contributions in units of $10^{-6}$ or ppm}
\label{table:syst}
\end{table}

For ${}^{12}$C and ${}^{208}$Pb, the longitudinal
polarization measurements included
data taken with a Compton polarimeter, yielding $P_{\rm e} = 0.8820 \pm 0.012\pm0.012$.
An independent M{\o}ller polarimeter gave
$P_{\rm e} = 0.9049 \pm 0.001\pm0.011$ for ${}^{12}$C and ${}^{208}$Pb.  
We used the average of these two measurements.
For the $^1$H and $^4$He data only
the M{\o}ller polarimeter was used. 
For $^1$H data, $P_{\rm e}=75.1\pm1.7$\% and for the $^4$He data, 
$P_{\rm e}=84.2\pm1.7$\%.

\begin{table}
\begin{tabular}{|c|c|c|c|c|}\hline
&&&&\\
Target    &       H &    $^4$He &     $^{12}$C &    $^{208}$Pb \\
\hline
$A_{\rm n}$(ppm)&$-6.80$&$-13.97$&$-6.49$ &$0.28$ \\
$\sigma(A_{\rm n})$(ppm)&$\pm1.54$&$\pm1.45$&$\pm0.38$ &$\pm0.25$ \\
$\sqrt{Q^2}$ (GeV) &  0.31 & 0.28 & \ \ 0.099\ \  & \ \ 0.094\ \ \\
$A/Z$ & 1.0 & 2.0 & \ \ 2.0\ \  & \ \ 2.53 \ \\
$\hat A_{\rm n}$ (ppm/GeV) & -21.9  &  -24.9   &  -32.8       &  +1.2  \\
$\sigma(\hat A_{\rm n})$(ppm/GeV)&$\pm5.0$&$\pm 2.6$&$\pm 1.9$ &$\pm 1.1$ \\
\hline 
\end{tabular}
\caption{The measured $A_{\rm n}$ and derived $\hat A_{\rm n}$ 
values (Eq. ~\ref{eq:An_hat}) for the four nuclei
along with the corresponding total uncertainties, $A/Z$ and $Q$.
}
\label{table:sum}
\end{table}

A summary of the systematic and statistical uncertainties is shown 
in Table~\ref{table:syst}.  The central values of $A_{\rm n}$ for each nucleus and
the total combined statistical and systematic uncertainties added in quadrature are displayed in the first
two rows of Table~\ref{table:sum}. 
%The results are as follows:
%\[A_n^H=-6.80\pm1.52{\rm (stat)}\pm0.27{\rm (syst)ppm},\]
%\[A_n^{He}=-13.97\pm1.39{\rm (stat)}\pm0.41{\rm (syst)ppm},\]
%\[A_n^{C}=-6.49\pm0.36{\rm (stat)}\pm0.10{\rm (syst)ppm},\]
%\[A_n^{Pb}=+0.28\pm0.21{\rm (stat)}\pm0.14{\rm (syst)ppm}.\]
For $^1$H, our result is consistent with a previously reported measurement~\cite{Maas:2004pd} 
for the same $Q^2$ but at a lower beam energy (0.85 GeV).
\begin{figure}
\includegraphics[width=1.0\columnwidth]{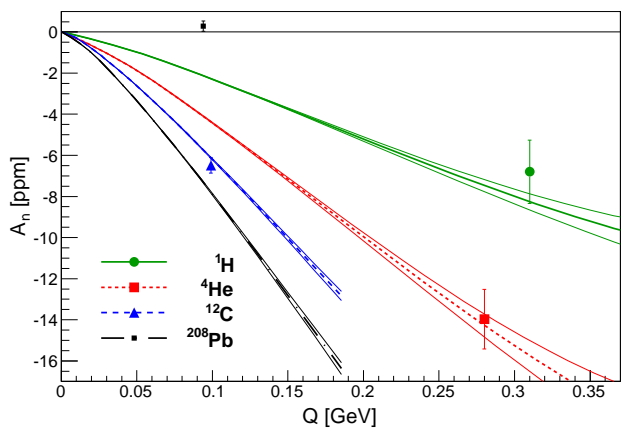}
\caption{Extracted physics asymmetries $A_{\rm n}$ vs. $Q$. Each curve, specific to a particular nucleus as indicated,  is a theoretical calculation from Ref.~\cite{Gorchtein:2008dy}. }
% The theoretical expectations from reference
% ~\cite{Gorchtein:2008dy} for $^{4}$He and $^{208}$Pb are shown as diamonds.}
\label{fig:An_vs_q}
\end{figure}

We now discuss the observed trends of the first ever measurements of $A_{\rm n}$ for target nuclei with $A>2$.
In our kinematic range, the calculations in ref.~\cite{Gorchtein:2008dy} scale approximately
with $Z$, $A$, and $\sqrt{Q^2}$ as
\begin{equation}
A_{\rm n}=\hat A_{\rm n} \frac{QA}{Z},
\label{eq:An_hat}
\end{equation}
where $\hat A_{\rm n}$ is approximately constant, with a small additional dependence on the incident beam energy
$E_{\mathrm beam}$: $\sim -25$ ppm/GeV for $E_{\mathrm beam} \sim 3$ GeV (for $^1$H and $^4$He), and
$\sim -30$ ppm/GeV for $E_{\mathrm beam}\sim 1$ GeV (for $^{12}$C and $^{208}$Pb). 
In the last two rows of Table~\ref{table:sum}, we see that the extracted $\hat A_{\rm n}$ from the three lower $Z$ nuclei
are consistent with this empirical trend, while the $^{208}$Pb result is consistent with zero. 
The $^{208}$Pb result is in strong disagreement with the theoretical prediction as shown 
in Fig.~\ref{fig:An_vs_q}, which plots the measurement results and their predictions \cite{Gorchtein:2008dy}.

\begin{figure}[tb]
\includegraphics[width=1.0\columnwidth]{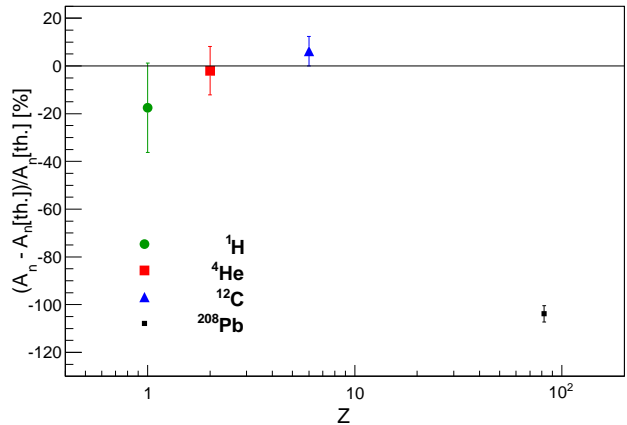}
\caption{\%\ fractional deviation of $A_{\rm n}$ measurements normalized to respective theory prediction vs. target nucleus $Z$.}
\label{fig:relAn_vs_Z}
\end{figure}

Motivated by this large observed disagreement, 
we initiate a discussion of the potential dynamics by first noting that the scattering
angle for all four measurements was roughly the same (Table~\ref{table:kin}). If dispersion corrections 
play a bigger role than predicted, one might expect larger disagreements for $A_{\rm n}$ measurements taken at lower beam energy, as is the case for $^{12}$C and $^{208}$Pb. However, the measured
$A_{\rm n}$ for $^{12}$C is quite consistent with theoretical expectations. 
In Fig.~\ref{fig:relAn_vs_Z}, we plot the fractional difference between the measured values and the 
predictions of Ref.~\cite{Gorchtein:2008dy} as a function of $Z$. 
The trend suggests that Coulomb distortions are playing a very significant role at large $Z$,  
underscoring the potential interest in 
additional $A_{\rm n}$ measurements with intermediate $Z$ nuclei.

 In conclusion, we have measured the beam-normal single-spin asymmetry $A_{\rm n}$ 
for $^1$H, $^4$He, $^{12}$C, and $^{208}$Pb and find good agreement for 
$^1$H, $^4$He and $^{12}$C with the calculations 
in ref.~\cite{Gorchtein:2008dy}, which include a dispersion
integral over intermediate excited states.  However, they are only to
order $\alpha^2$ (two-photon exchange) and neglect Coulomb distortions. On the other hand, 
$A_{\rm n}$ for $^{208}$Pb is measured to be very small and disagrees completely
with theoretical calculations. Coulomb distortions were shown in
ref.~\cite{Cooper:2005sk} to grow
rapidly with $Z$. 
On the other hand, the weight of dispersion corrections varies with the incident beam energy. Thus,
new theoretical calculations that treat dispersion corrections and Coulomb distortions simultaneously
as well as a systematic set of $A_{\rm n}$ measurements for a range of $Z$ at various beam energies might lead to new insights into the structure of heavy nuclei.

\begin{acknowledgments}
We wish to thank the entire staff of JLab for their efforts to develop and
maintain the polarized beam and the experimental apparatus. 
This work was supported by DOE contract DE-AC05-84ER40150 Modification No.~M175, 
under which the Southeastern Universities Research Association (SURA) 
operates JLab,
contract DE-AC02-06CH11357 for Argonne National Lab,
and by the Department of Energy, the National 
Science Foundation, the INFN (Italy), and the 
Commissariat \`a l'\'Energie Atomique (France).

\end{acknowledgments}

\end{document}